\documentclass[preprint2]{aastex}%

\shorttitle{Physical Characteristics of Accretion Induced Collapse}
\shortauthors{Taani et al.}

\begin{document}

\title{Investigation of Some Physical Properties of Accretion Induced Collapse in Producing Millisecond Pulsars}

\author{Ali Taani\altaffilmark{1} and ChengMin Zhang\altaffilmark{1}}
\affil{ National Astronomical Observatories, Chinese Academy of Sciences, Beijing 100012, China}
\email{alitaani@bao.ac.cn}

\author{Mashhoor Al-Wardat\altaffilmark{2}}
\affil{ Department of Physics, Al-Hussein Bin Talal University, P.O.Box 20,
71111, Ma'an, Jordan}
\and

\author{YongHeng Zhao\altaffilmark{1}}
\affil{ National Astronomical Observatories, Chinese Academy of Sciences, Beijing 100012, China}

\altaffiltext{1}{National Astronomical Observatories, Chinese Academy of Sciences, Beijing 100012, China}
\altaffiltext{2}{Sabbatical visitor, Physics Department, Yarmouk University, P.O.B. 566 Irbid, 21163 Jordan
}



\begin{abstract}
We investigate some physical characteristics of Millisecond Pulsar
(MSP) such as magnetic fields, spin periods and masses, that are
produced by Accretion Induced Collapse (AIC) of an accreting white
dwarf (WD) in stellar binary systems. We also investigate the
changes of these characteristics during the mass-transfer phase of
the system in its way to become a MSP. Our approach allows us to
follow the changes in magnetic fields and spin periods during the
conversion of WDs to MSPs via AIC process. We focus our attention
mainly on the massive binary WDs ($M \gtrsim 1.0 M_{\odot}$)
forming cataclysmic variables, that could potentially evolve to
reach Chandrasekhar
limit, thereafter they collapse and become MSPs. 
Knowledge about these parameters might be useful for further
modeling of the observed features of AIC.


\end{abstract}


\keywords{Stars: neutron stars, white dwarfs, cataclysmic variables, x-ray binaries, fundamental parameters}


\section{Introduction}
Cataclysmic variables (CVs) are short-period binary systems
consisting of massive white dwarfs (WDs) primary that is accreting
material via Roche-lobe overflow from low-mass secondary stars
(see i. e. \cite{1995CAS....28.....W}; \cite{2002ASPC..261..406W};
\cite{2011MNRAS.tmp.1594M}), so Accretion Induced Collapse (AIC)
is favorable in these systems (\cite{2004Sci...303....1143V}).
However, the AIC scenario has been proposed as an alternative
source of recycled pulsars sufficient to obviate the difficulties
with the standard model (\cite{1991PhR...203....1B};
\cite{2010MNRAS.402.1437H}; \cite{2012AN....333..1}). While other
authors questioned the viability of an AIC origin for the recycled
pulsars on theoretical and statistical grounds
(\cite{1990Natur.347..741R}; \cite{1995JApA...16..255V}),
\cite{2008AIPC..968..194F, 2008AIPC..968..188F} argued that the
AIC channel can form binary Millisecond Pulsars (MSPs) of all
observed types with $B \sim 10^{8-9} $G, $P \sim 20 $ ms and
$e\sim 0.44$. (e.g. \cite{1991PhR...203....1B};
\cite{2008LRR....11....8L}; \cite{2011A&A...526A..88W}). The
required conditions for the formation of an AIC in the case of a
steady accretion were summarized in \cite{1984ApJ...277..791N,
1987ApJ...322..206N}, \cite{1994MmSAI..65..339I} and in two recent
works \cite{2011MNRAS.410.1441C} and \cite{2012AN....333..1}.
However, unless the accretion rates are very high ($\rm>
10^{-7}M_{\odot}yr^{-1}$), the accreted (Hydrogen) matter will not
go into steady nuclear burning on the surface of the WD, and hence
its mass will never grow. In the situation of lower accretion
rates, the accreted matter will burn in Super Novae (SN) type Ia,
such that the WD will never reach Chandrasekhar limit
($\rm1.44~M_{\odot}$).



Nowadays, there is an increasing amount of evidence which points to
the fact that at least a fraction of low-mass X-ray binaries (LMXBs)
should be the result of AIC of a WD in a low-mass binary system
(\cite{1986ApJ...305..235T}; \cite{2010ScChG..53S...9L}). However,
this straightforward hypothesis had to face several problems, some
of which are not yet completely solved. The behavior of the external
layers of an accreting WD depends basically on the rate of mass
transfer and on the chemical composition of the accreted material.
Even though a lot of work has been done on this issue, much of it is
based on very restrictive assumptions (steady spherical accretion,
mass for instance).

In general, accretion of material lighter than carbon will induce
its burning. Only steady burning or non-disruptive flashing
regimes can be allowed in NS formation scenarios
\citep{2005MNRAS.356.1576W}. For which conditions this actually
happens is still largely undetermined, but it seems that there are
several combinations of the parameters mentioned above which allow
the star to retain the accreted material and to grow in mass. Even
though the WD manages to accrete mass steadily, the subsequent
collapse is not yet guaranteed because explosive thermonuclear
burning always happens in the contracting core before reaching
Chandrasekhar limit. The burning might propagate through the
entire star, releasing enough energy to blow it apart
\citep{2009JPhCS.172a2037W}.

Observable parameters of binary MSPs, e.g. mass of the pulsar, mass
of the companion, spin period of the pulsar, orbital period, orbital
eccentricity, etc., are used to probe the past accretion history of
the MSPs.

AIC produces WDs having mass less than Chandrasekhar limit,
because at least $\sim10\%$  of the mass accreted by the WD goes
to the binding energy of the nascent NS
(\cite{2011MNRAS.413L..47B}); this mechanism is similar to LMXB.
While long orbital periods could potentially occur via explosion
power from supernova type Ia, which, circumstances has no
observational counterparts, consequently, create single MSP. It is
currently unknown whether such AIC process appears and how much it
contributes to form MSPs population, this is a focus of much
present-day research.

The purpose of this paper is to demonstrate how to infer some of
the observable quantities and relate them to the dipolar magnetic
magnetic field at the surface of the NS, bottom field strength
(when the magnetospheric radius $R_{M}$ = R) and spin period,
during the AIC of a WD in its way to become a member of MSP
family.

To accomplish this goal, the model developed by
\cite{2009MNRAS.397.2208Z} is extended to include the magnetic
fields and radii for cataclysmic variables (CVs) observed sample.
Thus a significant reduction of the polar field strength could
occur if such a system accumulates the required critical mass
$M_{crit}$. Thus we demonstrate the observed properties of the
magnetic CVs, that can generally be explained by the model where
the field is at most only partially restructured due to accretion.

\begin{figure}[h]
\includegraphics[angle=0,width=8.0cm]{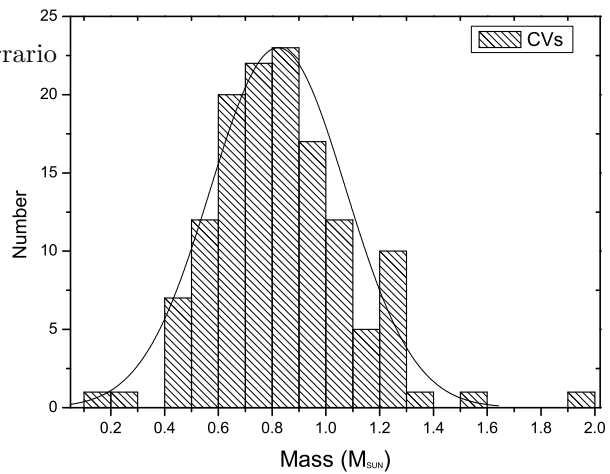}
\caption{Mass distribution of all CVs observed sample. The solid
line is the curve fitted using a Gaussian function. The data are
taken from Ritter \& Kolb (2011) to construct this distribution.}
\label{fig-0}
\end{figure}


\begin{figure}
\includegraphics[angle=0,width=9cm]{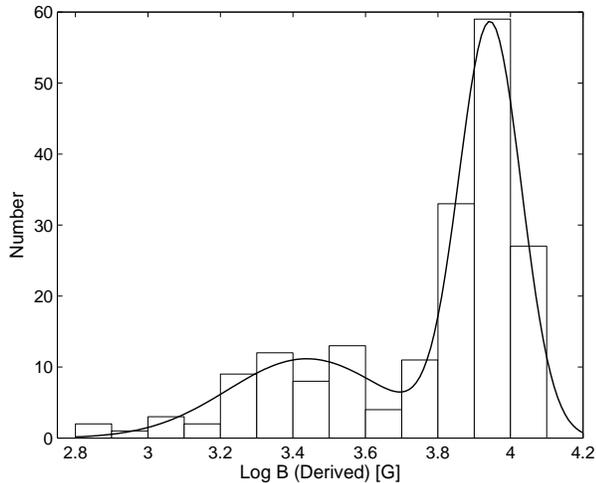}
\caption{Histograms of the bottom magnetic field in CVs. The
distribution is relatively Gaussian. Here the field strength of most
CVs has been inferred on assumption that CVs of non-spherical
accretion on to magnetized NS with a centered dipolar field (see
\cite{2009MNRAS.397.2208Z}). The solid line is the curve fitted
using a Gaussian function.} \label{fig-1}
\end{figure}

\begin{figure}
\includegraphics[angle=0,width=8.0cm]{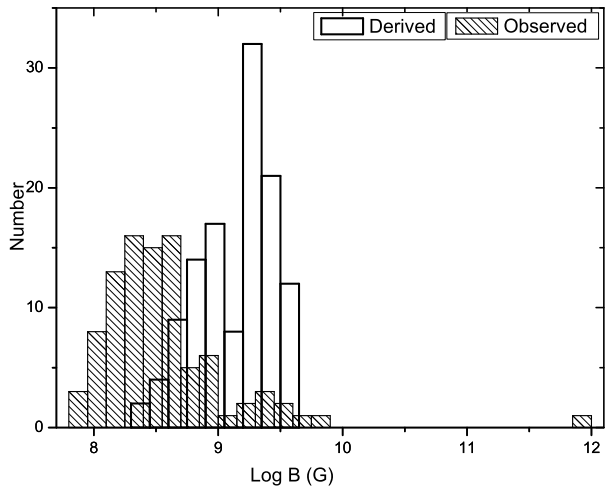}
\caption{Observed magnetic field distributions of MSPs  (shaded
histogram) and derived ones of the CVs  (solid histogram), which
were calculated by assuming the conservation of the magnetic flux
(see the text). The data for MSPs were taken from the ATNF catalogue
\citep{2005AJ....129.1993M}.} \label{fig-2}
\end{figure}


\section{Methodology}
\subsection{The Observed Population of CVs}

Most of the behavior of CVs can be understood in the context of a
close binary with mass transfer from a low-mass,
near-main-sequence companion star. The mass donor is known as the
secondary, and the star which accretes mass is known as the
primary. In CVs the primary is a WD, and the secondary is usually
a late-type roughly main-sequence star. Interacting binaries are
interesting and important to observe, as they allow the study of
accretion on observable timescales. Moreover, much can be learnt
about binary evolution from them, in particular the effects of
accretion on their ultimate evolution. It is predicted that above
the $2 - 3$ hr orbital period gap, angular momentum losses are
driven by magnetic braking (\cite{2003ApJ...599..516I}), while
below the gap the systems evolve via the emission of gravitational
wave radiation (see i.e. \cite{1983A&A...124..267S};
\cite{2011ApJS..194...28K}).
However, the theoretical relationship between the mass and radius of
a WD is essential in interpreting some  observational results. M-R
relation was first defined by \cite{1939isss.book.....C} as:

\begin{equation}
 R \propto \frac{1}{ M^{1/3}}
\end{equation}

 \noindent for Sirius B (${\rm~M=0.986M_{\odot}}$, ${\rm~R=0.00846R_{\odot}}$),
 where mass and radius are measured in standard solar radii and
 masses (\cite{1995LNP...443...41W}).
We use this M-R relation to derive the radii of CVs which are not
yet observed. The results of the calculated M-R values for massive
CVs are listed in  Table~\ref{tbl-1}.

The sample of mass of CVs we have considered is the set of binary
systems collected by \cite{2011yCat....102018R} from several
high-quality surveys. The sample consists then of 133 systems.
Fig. 1  shows the Gaussian distribution of mass of CVs, with mean
at ${\rm~M_{CV}\sim 0.822M_\odot}$, where the median mass of these
sources is ${\rm~M_{CV} = 0.81\pm 0.16M_\odot}$. Since the  Ritter
and Kolb (2011) catalog gives a mass of CH UMa ${\rm~M_{CV}\sim
1.9\pm0.30M\odot}$ (well above the Chandraskhar mass limit for a
WD), we decided to exclude this source from our study. We fit the
Gauss function to the mass distributions. The Gauss function we
choose reads,
\begin{equation}
y = y_{0} + \frac{A}{w_{0} \sqrt{\frac{\pi}{2}}} \, \exp \left(-2
\left(\frac{x-x_0}{w_{0}} \right)^2 \right)
\end{equation}
where $y_{0}$, $x_{0}$, $w_{0}$ and $\emph{A}$ are offset of y-axis,
center of x-axis, width and area represented by the curve
respectively. The fitting results are listed in Table 2.








\subsection{Magnetic field}
The strength of the magnetic field influences the physical
conditions in the accretion column, and as a consequence the
dominate radiation mechanisms and
 the evolutionary phase of CVs population (see i.e. \cite{1983A&A...124..267S};
\cite{1994MNRAS.269..907K}; \cite{1983ApJ...275..713R};
\cite{2011MNRAS.414L..16Q}). To derive the bottom magnetic field
in CVs (where the field is at most only partially restructured due
to accretion), we follow \cite{2009MNRAS.397.2208Z} model, which
suggest that the bottom filed $B_{f}$ for WDs is reached when the
magnetospheric radius $R_{\rm M}=R$, giving
\begin{equation}
B_{f,WD} = 2.8 \times 10^{3} \, \dot{M_{16}}^{1/2} \, m^{1/4} \,
R_{9}^{-5/4} \, \phi^{-7/4} \, ~{\rm (G)}
\end{equation}

\noindent where, $m = M/M_{\odot}$ is the WD's mass in CVs $M$,
where the mean mass of these sources is ${\rm~M_{CV} =
0.822M\odot}$, $ R_9$ is the WD radius in units of $10^9$~cm,
$\phi$ is a parameter which is estimated to be $\sim 0.5$ for disc
accretion, and $\dot{M}_{16}$ is the accretion rate in units of
$10^{16}$ gs$^{-1}$. The concept is displayed pictorially in
Fig.~\ref{fig-1}, which shows clearly two peaks (bi-modal)
Gaussian distributions for most of the CVs with magnetic fields in
the range (0.60 - 11.20)$\times10^{3}$ G. We fit the Gauss
function (same as Eq. (2)) to this distribution.  A full summary
of the main fitting results  is given in Table 3.


Another interesting point is to investigate the correlation between
the magnetic fields of MSPs with bottom fields of CVs, which can be
derived according to the conservation of the magnetic flux, in
other words, study the magnetic fields that produced by AIC. Let's
start, then, with the conservation of the magnetic flux

\begin{equation}
B_{NS} = B_{f,WD} \times \left(\frac{R_{WD}}{R_{NS}}\right)^2
\end{equation}

As an example of a typical NS we suggest to adopt a radius
($R_{NS}$) of about $15\times10^{5}$~cm, where $\alpha$ can be
expresses as $(R_{WD}/R_{NS})$, subsequently $\alpha^{2}\sim10^{5.5-6}$, and $B_{f}\sim10^3$G in CVs. As a result, one arrives
at an expression for the minimum value of magnetic field in NSs equals to:
\begin{equation}
B_{NS}\sim10^{8.5-9}  G
\end{equation}
which is somehow similar to the observational values of MSPs.

To study the contribution of magnetic field of CVs on the MSPs
population, we produced a histogram of the total sample of MSPs
(Fig. 3), which shows a clear Gaussian distribution for most of
the objects, while the observed sample shows two peaks and a small
fraction of these systems have intersection. We note that the
pinnacle of derived histogram is much higher than in observed,
this is due to the sensitivity of the radio observation process
for MSPs. Since all pulsar surveys have some limiting flux density
(\cite{2008LRR....11....8L}). However, new data acquisition
systems provide significantly improved sensitivity to MSPs and are
being used to resurvey the sky. Therefore the observed numbers of
MSPs of magnetic field are relatively high compare with CVs
population. It is noteworthy to mention here that we also found
that $\sim 15\%$ of MSPs could potentially be formed from CVs via
AIC process. The peak of the distribution for derived sample
occurs at $\sim \times10^{9.4}$G comparing with $\sim
\times10^{8.5}$G for the observed one. Furthermore, very high
magnetic field Polar systems (like AM Her) are not represented in
this figure.


\subsection{Spin period}
It is well established that the NS in binary MSP systems forms
first, descending from the initially more massive of the two binary
stellar components. The NS is subsequently spun-up to a high spin
frequency via accretion of mass and angular momentum once the
secondary star evolves (For a more detailed review, see i.e.
\cite{1982Natur.300..728A}; \cite{1991PhR...203....1B};
\cite{2000ApJ...530L..93T}; \cite{2011MNRAS.416.2285L};
\cite{2011arXiv1106.0897T}) In this section we will determine the
spin period of MSPs originated from WDs, the process begins with
simple Keplerian frequency, the angular velocity of the NS is equal
to the Keplerian angular velocity of the magnetosphere, at roughly
the Alfv$\acute{\textrm{e}}$n surface,


\begin{equation}
v_{K} \propto R_{NS}^{-3/2} \rightarrow P_{MSP} \sim R_{NS}^{3/2}
\end{equation}
from which we obtain $P_{MSP}$ as a function of $P_{WD}$
\begin{equation}
P_{MSP} \sim P_{WD, min} \left(\frac{R_{NS}}{R_{WD}} \right)^{3/2}
\end{equation}

 where the minimum spin period of the standard WD (\cite{1994MNRAS.267..577D}; \cite{ 2011MNRAS.tmp.1594M})
 taking value $\sim 30$ s, $R_{NS} = 10$ km and
$R_{WD} = 1000 $ km (see Eqn 1) , giving
\begin{eqnarray}
  P_{MSP} \sim 1 ms \nonumber \end{eqnarray}

Note that the estimated spin periods are
not  far from the true spin periods of the MSPs. This clearly
supports the view that the AIC effect play an important role
in deriving  spin periods for MSPs.

Fig. 4 shows the observed and derived spin period distributions of
MSPs. It can be found that most of derived values centered at 0.011
s and ranging from 0.011 to 52.8 s. As for the observed MSPs,  shows
relatively Gaussian and regular distributions for most objects,
centered at 4.4 ms and ranging from 1.4 ms to 19.5 s. According to
these distributions, the ratio of MSPs originated from CVs is about
$\sim 10\%$. This result agrees with some theoretical predictions
such as those by \cite{1995CAS....28.....W} and
\cite{2002ASPC..261..406W}.


\begin{figure}[h]
\includegraphics[angle=0,width=8.0cm]{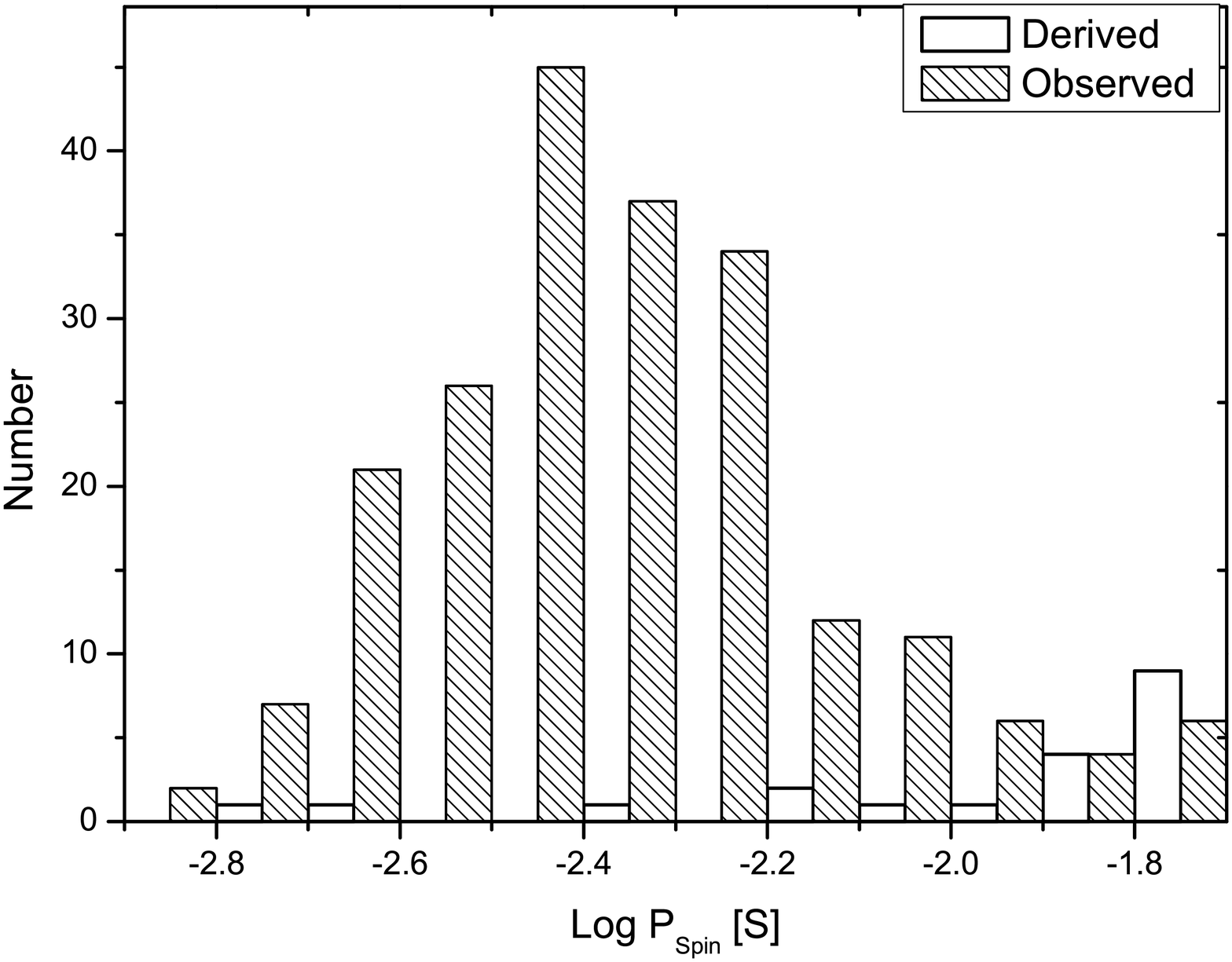}
\caption{The distribution of MSPs on the base of spin-period;
Observed (shaded histogram) and  derived (solid histogram). The data
for MSPs were taken from the ATNF catalogue
\citep{2005AJ....129.1993M}} \label{fig-4}
\end{figure}





\subsection{Mass}
Most of the stellar binary systems consist of two well-separated
components, with a little or no interaction beyond their mutual
gravitational attraction. However, in close binary systems tidal
effects become important, leading to a circularization of the orbit,
and spinning up the star's rotational speeds so that they become
synchronous with the orbital motion.

The  current edition
(7.16)\footnote{http://www.mpa-garching.mpg.de/RKcat/} of the
Catalogue of Cataclysmic Binaries \cite{2011yCat....102018R}
contains a wide variety and natures of sources such as 880 CVs, 98
LMXBs, and 319 related objects. Among them we have just 18 massive
CVs in the range ${\rm~1.1 - 1.3M_{\odot}}$, which is considered a
relatively small number. A summary of the known properties of
these systems is given in table~\ref{tbl-1}). If matter is
accreted at a rate of ${\rm\dot{M}\sim10^{-9}M_{\odot}/yr ~to~
10^{-8}M_{\odot}/yr}$ (\cite{2006MNRAS.366..137Z}), and the total
mass accreted exceeds a critical value $\rm\Delta M_{crit}\sim
0.1- 0.2M_{\odot}$ with $10^9$ yrs, then the massive CV will be
recycled to become a MSP after the mass reaches the Chandrasekhar
limit.  It is worth noting here that the mass of the newly formed
object will be below the maximum mass for MSP
(\cite{2004Sci...303....1143V}).
 This provides evidence for the AIC scenario in massive CVs and
evolutionary hypotheses of MSP birthrate. In sum up, whether a CV
might possibly do this will depends on: (i) the mass of the WD in
the system, and  (ii)  the mass transfer rate that the donor star
may supply, because we know that at low mass transfer rates one only
gets SN explosions Type Ia, and the WD will never grow to become a
NS.

It should be noticed that the AIC process leads to MSP with mass
less than Chandrasekhar limit. When a WD gains mass by accreting
matter from its companion, its binding energy also increases, this
effect is significant in the estimation of the amount of mass
accretion.
 However, unfortunately most of the studies argued that the binding
energy of the NS is not commonly considered
(\cite{2011MNRAS.413L..47B}).












\begin{table*}
\begin{center}
\caption{List of physical properties for the all massive CVs ($M
\geq 1.0 M_{\odot}$) \tablenotemark{*} \label{tbl-1}}

\begin{tabular}{lcccl}
\tableline\tableline
Name   &     $M/M_{\odot}$  &   $P_{Orb}$ (D)   &  $R_{Derived} ({10^9}$cm)    &   $B_{Derievd} ({10^3}$G) \\
\tableline
CAL 83  &   1.3 &   1.04    &   0.76    &   13.95  \\
CI Aql  & 1.2 &   0.618   & 0.83    &   12.38    \\
BV Cen  &   1.24 &   0.61 &   0.8 &   13  \\
EY Cyg  &   1.1 & 0.49    &   0.9 & 10.86    \\
RU Peg  &   1.21 & 0.37    &   0.82 &   12.53    \\
QZ Aur  & 1.05 &   0.357 &   0.95 &   10.13    \\
MU Cen  &   1.2 & 0.342   &   0.83 & 12.38   \\
BF Eri  &   1.28 & 0.27    & 0.78 & 13.63    \\
VY Scl  &   1.22 & 0.232 &   0.81    & 12.69    \\
RX And & 1.14 & 0.209   & 0.87    & 11.46    \\
SS Aur  & 1.08    & 0.182   &   0.92 & 10.57    \\
BD Pav  & 1.15 & 0.179   & 0.86    & 11.61    \\
U Gem & 1.2 & 0.176 & 0.83 & 12.38  \\
V603 Aql    & 1.2 & 0.138 & 0.83 & 12.38    \\
WW Hor & 1.1 & 0.08 & 0.9 & 10.86  \\
CU Vel & 1.23 & 0.078   & 0.81 & 12.84  \\
DP Leo  & 1.2 & 0.062 & 0.83 &   12.38 \\
\tableline \tablenotetext{*}{The masses and orbital periods were
taken from \cite{2011yCat....102018R}, while B and R are
determined by the condition that the magnetosphere equal to the WD
radius (Zhang et al. 2009). The model-dependent parameters are
$\phi = 0.5$ and $\dot{M} = 10^{16} \rm gs^{-1}$, (see the text).}
\end{tabular}
\end{center}
\end{table*}

\begin{table*}
\centering
 \caption{\bf The fitting results of distributions for mass of Cvs.}
\setlength{\tabcolsep}{3pt} \centering \label{table1}
\begin{tabular}{lccccccl}
\hline \hline
quantity  &   $y_{0}$ &  $x_{c}$  & $w$ & \emph{A}  & $\sigma$ & $\chi^2$/Dof & $R^2$ \\

\hline
\\
M$_{CVs}$  &   $0.53 \pm0.64$&$0.79 \pm0.012$ & $0.40 \pm0.03 $ &
$12.30 \pm0.92 $ & 0.20 & 3.311
 & 0.955

  \\
\hline
  \hline\\

\multicolumn{4}{l}{}
\end{tabular}
\medskip
\end{table*}

\begin{table*}
\centering
 \caption{\bf The fitting results of distributions for derived magnetic field strength in CVs .}
\setlength{\tabcolsep}{3pt} \centering \label{table1}
\begin{tabular}{lcccccccl}
\hline \hline
quantity  &   $y_{0}$ &  $x_{0}$  & $x_{1}$& $w_{0}$ & $w_{0}$& \emph{A} & \emph{B} &  $R^2$ \\

\hline
\\
 B$_{Derived}$ &   $-0.05\pm0.004 $& $3.93\pm0.04 $ & $3.41\pm0.08 $& $0.186\pm0.008 $
 & $0.37\pm0.003$& $14.42 \pm4.57 $ & $5.45\pm1.35 $& 0.942
 \\





%
\hline \hline\\

\multicolumn{4}{l}{}
\end{tabular}
\medskip
\end{table*}

\subsection{Orbital periods and eccentricity}

We now believe that in the case of spherical SN explosions in which
C-O WD (at low mass transfer rates) produced impart significant
kicks to the companion that make their survival prospects within
binary systems in a long orbital period or bleak.
The asymmetric mass loss during the AIC process is expected to
provide kick velocities of about 50 ${\rm km s^{-1}}$ or less
(\cite{2000ApJ...531..345G}).
 Consequently, the AIC produces very low
eccentricities ($e < 0.1$). When the natal MSPs receive a
relatively strong kick ($\rm> 100 km s^{-1}$),  the AIC channel
produces eccentric binary MSPs in the Galaxy. Such a kick seems to
be highly unlikely in the AIC process, hence the probability of
forming eccentric binary MSPs via the AIC channel can be ruled
out. Even if a high kick is allowed, the AIC channel cannot
produce eccentric with an orbital period of $\rm\geq70$ d
\cite{2011MNRAS.410.1441C}.


In sum up, the high-mass binary systems tend to have highly
eccentric orbits, while the low-mass binaries have smaller
eccentricities.
The origin of MSPs in low-mass binaries could be due to either AIC
or supernova explosion. If MSPs are formed by the AIC process,
their average velocity would be much lower than that of normal
pulsars. Further work will certainly clarify this issue as known
sources are better characterized.

\section{Summary and Conclusions}

We have studied the possible AIC scenario in massive CVs, through
simple calculations of some physical characteristics such as
magnetic field, spin period and mass. The basic information that
goes into the calculation can be summarized as follows

\begin{enumerate}
\item CVs would be invoked via the capability of producing a
significant portion of the MSPs population through the AIC
process, a regime which otherwise may be unattainable by the
normal channels that produce NSs (i. e. long orbital periods for
half of the MSPs and the formation of single MSPs) in order to
reduce the gap between standard LMXB evolution and the observed
population of MSPs.

\item We find that the quantitative implications of our
calculations are that, we can estimate the expected magnetic field
strength in the observed population of MSPs which could be
contributed from CVs is about  $\sim 15\%$. While it turns out that
the contribution in spin period is about $\sim 10\%$. It would be
very interesting though, if observations would yield either a very
slowly spinning (few hundred ms) pulsar associated with a very low
magnetic field, or a high magnetic field MSP with a very rapid spin.
(However, the latter kind of pulsars might be unlikely to be
detected given that its very strong magnetic dipole radiation would
slow down its spin rate within a few Myrs).

\item We compute the bottom magnetic field in CV samples, with a
discussion on the implications of the model in
\cite{2009MNRAS.397.2208Z}. We extend this model to conditions
appropriate to accreting WDs. Thus the model predicts that the
magnetic polar cap widens as material is accreted with the field
being advected towards the equatorial regions by the ensuing
hydrodynamic flow.

\item On the basis of our results we conclude that, it is possible to
make comparisons about the relative importance of spin periods,
magnetic field and some other parameters in CV population, that
potentially produced via AIC process about their aggregate impact in
the observed population of MSPs.

\item The mass of MSPs which produced via AIC process is less than
Chandrasekhar limit, because at least $\sim 10\%$ of the original
mass goes to the binding energy of the NS.

\item We further find that the predictions of some parameters
after AIC process for the average levels are consistent with the
observed population (such as spin period, magnetic field and
mass). Future work will go steps further, and consider other
observable quantities such as orbital period, eccentricity and
mass ratio (q) for MSP populations using the more extensive data
set now available from \cite{2011yCat....102018R} and other
sources.

\end{enumerate}







Ultimately, the critically unique analysis will be the cross
correlation with the CVs, AIC process and MSPs, which is well
beyond the scope of this letter. Our main conclusion is that the
simple calculations presented herein, based on the standard
relationships between mass, magnetic field, spin and orbital
periods, required to support the more sophisticated treatments of
the problem, and suggests that how the AIC contributes to the
observed population of MSPs.





\section*{Acknowledgements}
The research presented here has made extensive use of the 2012
version of the ATNF Pulsar Catalogue (Manchester et al. 2005). Dr.
Al-Wardat would like to thank Max Planck Institute for
Astrophysics-Garching for the hosting funded by the Deutsche
Forschungsgemeinschaft (DFG)  during the writing of this work. This
research has been supported by NSFC (No.10773034)  and National
Basic Research Program of China (2012CB821800). We thank the
anonymous referee for a careful reading of our manuscript and for
numerous useful comments.

\end{document}